\documentclass{aastex}

\shorttitle{The 3.3 $\mu$m Emission Feature and UV Excitation}

\begin{document}

\title{Does the 3.3 $\mu$m PAH Emission Feature Require UV Excitation?}
\author{Tracy L.\ Smith\altaffilmark{1},
\& Geoffrey C.\ Clayton\altaffilmark{2}
\& Lynne Valencic\altaffilmark{3}}
\altaffiltext{1}{Space Science Institute, 3100 Marine Street, Ste A353,
Boulder, CO 80303-1058, tsmith@campbell.mps.ohio-state.edu}
\altaffiltext{2}{Department of Physics \& Astronomy, Louisiana State
University, Baton Rouge, LA  70803, gclayton@fenway.phys.lsu.edu}
\altaffiltext{3}{Instituto de Astronomia, Universidad Nacional Autnoma de Mexico,
PO Box 439027, San Diego, CA  92143, valencic@astrosen.unam.mx}

\begin{abstract}

The unidentified infrared bands (UIBs) have been observed in virtually every dusty astrophysical 
environment investigated.  The UIB carrier must be abundant and ubiquitous. 
Strong evidence 
points to polycyclic aromatic hydrocarbons (PAHs) as likely candidates, but 
the identification is not 
complete.  Additional diagnostics are needed to further constrain the UIB carrier, such as probing 
excitation sources which range from UV-strong to UV-weak, in order to determine the ``band gap'' of 
the UIB carrier.  Observations and models suggest that the UIBs can be found in sources with 
weak UV fields.  To 
that end, we present new results of observing the 3.3 $\mu$m spectral region in six stars 
embedded in reflection nebulae, and in six Vega-like stars.  These objects 
have effective temperatures ranging from  3500 to 12,000 K. 
Their environments include dust that should be relatively unprocessed (reflection nebulae) and dust 
that has most likely undergone significant processing (Vega-like) by the embedded illumination source.  
Together with data from the literature, we have a sample of 27 sightlines. Our analysis suggests 
that neither the strength of the UV field impinging on the dust nor the effective temperature of the 
star is the determining factor in whether the 3.3 $\micron$ UIB emission is present in an object.
We found three detections of the 3.3 \micron~emission band, one in a Vega-type object, one in a 
Herbig Ae/Be object and one in a reflection nebula, and all with disks.  The role of disk geometry 
is likely to be important in revealing or obscuring the photo-dissociation regions from which the UIB 
emission arises.
\end{abstract}


\section{Introduction}

The unidentified infrared bands (UIBs), at 3.3, 6.2, 7.7, 8.6, 11.3, and 12.7 $\mu$m, are 
the most prominent features in the near-infrared (NIR). They appear in spectra of the Galactic disk 
and high-latitude cirrus clouds (Matilla et al. 1996, Lemke et al. 1998) and other 
galaxies, including starburst galaxies (Roche 1989, Genzel and Cesarsky 2000), and in a wide range 
of astrophysical objects such as reflection nebulae (NGC 7023), planetary 
nebulae, dark clouds and photo-dissociation regions such as the Orion Nebula (Sellgren et al. 1990).  
Discovered in the 1970s (Gillett et al. 1973, Soifer et al. 1976 and Russell et al. 1977), today they 
have been observed to be present in most UV-rich astrophysical environments.  The proposed 
carriers of the UIBs include polycylic aromatic hydrocarbon (PAH) molecules (Leger \& Puget 1984; 
Allamandola, Tielens, \& Barker 1985) and grains containing PAHs (Duley \& Williams 1981, Leger 
\& Puget 1984).  Previous observations show UIBs occur in environments with moderate to high 
UV-excitation conditions, however, few UV-poor objects have been observed in order to determine 
the lower-limit for excitation of the UIB at 3.3 $\mu$m.  

UIB spectra can be split into two groups, those seen in 
interstellar dust (A-type) and those which are observed in circumstellar dust 
(B-type) (Geballe 1997; Tokunaga 1997).  
Class A-type UIBs, which include somewhat narrow, emission features at 3.3, 6.2, 7.7, 8.7, 11.3, and 
12.7 $\mu$m, are often observed in UV-rich sources.  Most UIB spectra appear to fall under Class A, 
having been observed in reflection nebulae, planetary nebulae, H II regions, starburst galaxies and 
even the diffuse interstellar medium (ISM).  Class B-type UIBs, which include narrow emission features 
at 3.3, 6.2, and 6.9 $\mu$m and broad emission bumps at 3.4, 8, and 12 $\mu$m (Geballe 1997; Tokunaga 1997), 
have been found in circumstellar dust illuminated by F- and G-type post-asymptotic giant branch (AGB) stars 
forming planetary nebulae (Uchida et al. 2000; Geballe et al. 1992).  In addition to spectra in these classes, 
another very small group of objects exhibits a unique pattern of features between 3.3 and 4 $\mu$m.  
Van Kerckhoven et al. (2002), observed 
features at 3.3, 3.43 and 3.53 $\mu$m in the star Elias 1 (A0-6), embedded in the reflection nebula IC 359 in 
the Taurus-Auriga complex (Elias 1978).  They attributed the 3.53 $\micron$ feature to the presence of 
nanodiamonds in the circumstellar material.  This unusual grouping has only been observed in two other 
objects, HD 97048, an A0pshe Herbig Ae/Be, and HR 4049, a B9.5 Ib-II star evolving into a protoplanetary nebula.

The central question, which motivated our observations, was whether the 3.3 $\mu$m UIB can be detected in 
weak UV field environments illuminated by stars with low T$_{eff}$.  This relates directly to the 
UV-absorption characteristics of the UIB carrier.  Models of PAH emission, building from extensive PAH 
laboratory work, suggest that a wide variety of PAHs may exist in dusty environments, with a typical 
size of 20-50 C atoms (Geballe et al 1989) (possibly larger), and absorption cross-sections dependent 
on size, molecular weight, and whether they are free-flying or incorporated into grains 
(Allamandola et al. 1989; Tielens 1993, Schutte et al. 1993, Cook and Saykally 1998; Bakes et al. 2001; 
Li \& Draine 2001).  No firm UV-absorption cutoff has been established for astronomical PAHs, nor has 
observational evidence of PAH absorption in the UV been found (Clayton et al. 2003), but laboratory 
evidence suggests little absorption in the visible (Sellgren 2001).  One might expect, therefore, to 
observe a relationship between the presence of the 3.3 $\mu$m UIB and environment (interstellar versus 
circumstellar), and specifically, with respect to the UV radiation field illuminating the dust.

It has been suggested that strong UV fields leave interstellar PAHs mostly neutral and positively charged, 
but weak UV fields are predicted to leave PAHs neutral or negatively charged (Bakes \& Tielens 1994, 1998; 
Salama et al. 1996; Dartois \& d'Hendecourt 1997).  Variations in the width of the 3.3 $\mu$m feature have been 
observed among different objects such as in Orion (Sellgren, Tokunaga, \& Nakada 1990), the reflection nebula 
NGC 7023, the planetary nebula (with a WC nucleus) IRAS 21282+5050 (Nagata et al. 1988) and toward the 
central star of the bipolar nebula HD 44179.  However, Sellgren et al. (1990) conclude that no dependence 
of width on UV field is evident.

Sellgren, Werner \& Allamandola (1996) found the equivalent width of the 3.3 $\micron$ UIB in reflection 
nebulae to be independent of $T_{eff}$.  Their coolest 3.3 $\micron$ detections were for the nebula vdB 10 
around HD 20041 (an A0 Ia-type star, $T_{eff}$ = 9400 K) and vdB 133 around HD 195593 (A+B) (an F5 Iab and 
a B7 II-binary, $T_{eff}$ = 6800 + 12,000 K).  Uchida et al. (2000) saw no spectroscopic differences among 
UIBs observed from 5-15 $\mu$m for sources with T$_{eff}$ ranging from 3600-19,000 K.  The binary HD 195593 
(A+B) illuminating vdB 133, is thought to be unique.  Uchida et al. (1998) note that vdB 133 is illuminated 
by the lowest ratio of UV to total flux of any nebulae observed to have the UIBs above 5 $\mu$m.  This led Li 
\& Draine (2002) to suggest that UV photons are not required to excite UIB emission. See Figure 1.

In this paper, we present new observations of vdB 133 and 11 other objects, chosen to cover both Class A and 
B-type sources with stars ranging from 3500-11,000 K in order to probe the dependence of the 3.3 $\mu$m emission 
feature on UV radiation field.  The following section (2) describes the data reduction, and in section (3), we
discuss the results and present our conclusions.

\section{Observations and Data Reduction}

The sample stars, listed in Table 1 were observed from 1.9 - 4.2 $\mu$m with SpeX (Rayner et al. 2003) at the 
3.0 m NASA Infrared Telescope Facility (IRTF) on Mauna Kea on 2002, August 28-30 and September 2.  Spectra of 
twelve targets were obtained using SpeX in the cross-dispersed mode using a 0.8$^{''}$ x 15 $^{''}$ slit size, 
oriented north-south, with a spectral resolution of R $\approx$ 950.  The data were obtained by nodding 
the telescope in the pattern on-target, off, off, on-target, with a typical total integration time of 100 
s acquired using two cycles and five co-adds.  Each target was observed three times, each bracketed by 
observations of a G-type main sequence star in order to accurately remove Telluric lines.  For each pointing, 
flats were taken using an IR lamp (1200 K) and arcs with an argon lamp, to be used later for wavelength calibration.

Data reduction of the target and G stars was accomplished through the use of the package SPEXTOOL (Cushing et al. 
2004; Vacca et al. 2003).  It was used to remove the sky background, calibrate, and extract the aperture of 
each on-off object pair, after which the pairs were summed together.  Individual observations of the same object 
were co-added.  The target spectra were then divided by the G stars observed at the same airmass and the result 
multiplied by a blackbody of appropriate temperature so that the absolute flux level of the target is preserved. 
The reduced spectra are plotted in Figure 2. 

Of the detections in our sample, two are mid A-type stars, Elias 1 and HD 169142, and one is classified as a G0 V 
star, HD 34700.  Van Kerckhoven et al. (2002) observed Elias 1 and Kraemer et al. (2002) observed HD 34700, 
both using the ISO-SWS.  Meeus et al. (2001), observed  HD 169142 and 13 other Herbig Ae/Be objects from 2-45 $\mu$m using 
the ISO Short Wavelength Spectrometer.  They found the 3.3 $\mu$m feature in six of the 
fourteen objects observed, including HD 169142(A5 Ve), HD 100546 (B9 Ve), HD 179218 (B9e), HD 142666 (A8 Ve), 
HD 100453 (A9 Ve), and HD 142666 (A8 Ve), ranging in T$_{eff}$ from 6250 to 11,000 K.  We have added the 14 stars 
in the Meeus sample to the twelve stars observed here, plus HD 20041 from Sellgren at al. (1996). There is one 
star (HD 169142) in common between the samples. They are listed in Table 2.  In addition to the 3.3 $\mu$m emission, 
one object, Elias 1, shows emission features at 3.43 and 3.53 $\mu$m, attributed to H-terminated diamond surfaces 
(Guillois et al. (1999); Van Kerckhoven, Tielens \& Waelkens 2002).

The illumination of vdB 133 by HD 195593 (A+B) is very interesting with respect to the subject of 
UIB excitation.  In spite of the presence of a B7 II star in the binary, it can be shown (Kurucz 1992) 
to emit only $\approx$ 16\% of its total luminosity shortward of 240 nm (Uchida et al. 2000) and 
$\approx$ 21\% shortward of 400 nm.  We did not detect the 3.3 $\micron$ UIB feature in vdB 133 
within $15^{''}$ NS of the star using IRTF, and it was not detected by Vandenbussche et al. (2002) using 
the ISO-SWS to create an atlas of near-infrared stellar spectra, Fig. 2.  However, Sellgren, Werner, \& 
Allamandola (1996) detected the 3.3 $\micron$ UIB at the star using a $19.6^{''}$ diameter aperture and indicated 
a weak feature detection at an offset of $30^{''}$E $30^{''}$W from the central star using the same aperture.    
Li \& Draine (2002) predict that the 3.3 $\mu$m UIB should be seen in vdB 133. This is shown in Figure 1. 
They suggest that even in regions ``devoid'' of UV photons, a PAH with 100 carbon atoms should be excited by a 
500 nm photon and emit strongly in the UIR bands  between 6.2 and 11.3 $\mu$m.  Uchida et al. (2000) detected 
the UIB features at 6.2, 7.7, 8.7, 11.3, and 12.7 $\micron$ at an offset of $100^{''}$ from vdb 133.

\section{Discussion}

Li \& Draine (2002) have suggested that UV photons are not required to excite UIB emission. The T$_{eff}$ of the 
exciting star will determine how much UV radiation is produced.  But the T$_{eff}$ does not take into account 
the dilution effect of the distance of the dust from the star.  A more useful parameter is the UV radiation field 
experienced by the dust at some distance from the star.  We estimated the UV radiation fields for the 27 objects
in our sample by integrating Kurucz (1992) stellar atmospheric models from 91-240 nm, and then multiplying by a 
dilution factor of $(R_{*}/d)^{2}$, where $R_{*}$ is the stellar radius and d is the distance from the star at 
which the UV radiation field is being estimated.  A UV-cutoff of 240 nm was chosen in order to make a conservative 
estimate of photons available for the excitation of the 3.3 $\mu$m UIB feature, as the UV-absorption spectrum for 
potential carriers such as PAHs is as yet unknown.  The UV radiation field $15^{''}$ NS of each star, corresponding 
to the slit length available on SpeX, is expressed as $G_{0}$. This is the local interstellar radiation field as determined 
by Habing (1968), in units of 1.6 x $10^{-3} erg s^{-1} cm^{-2}$ (Boulanger 1998, Dartois \& d'Hendecourt 1997, 
Uchida et al. 2000).  The estimated values of $G_{0}$ for each object are listed in Table 2. The $G_{0}$ value of 
2700 estimated for HD 195593 (A+B) at an offset of $15^{''}$, is consistent with the $G_{0}$ value from Uchida et al. 
(2000) of 75 for an offset of $100^{''}$ from the star.  Li \& Draine (2002) estimate a $G_{0}$ value of $\approx$ 1215 
``at the nebula'', using the linear combination of the Kurucz (1979) model atmospheres integrated for each 
star (A \& B) between 91.2-1000 nm.  

Figure 3 plots the estimated UV radiation field versus effective temperature of the exciting stars for our sample with
detections of the 3.3 $\micron$ feature marked. Clearly, no correlation exists between the strength of the UV 
radiation field ($G_{0}$) and the detection of the 3.3 $\micron$ emission feature in the observed sample (ours and 
those of Meeus et al.).  The calculation of $G_{0}$ assumes that the dust is situated at a distance from the exciting 
star corresponding to 15 $\arcsec$, projected on the plane of the sky.  Since matter around the stars is in many cases 
distributed out to 15 $\arcsec$ (the available SpeX slit length), this represents the minimum value for $G_{0}$ 
impinging on the dust.  Table 2 includes the IRAS total surface brightness of the sky surrounding a target in a 0.5 
degree beam at 100 $\mu$m; a value above 30 MJy/Sr indicating dust with appreciable column density.  As can be seen from 
the table, all but five of the stars for which measurements were available have surface brightnesses larger than 
30 MJy/Sr. The interesting exception being that for HD 34700, one of the three positive 3.3 $\mu$m emission detections 
in our original sample, which has a measurement of 28 MJy/Sr.

Since all the stars with detections in our sample have disks, the emission is likely to be mostly
from dust in a circumstellar disk much closer to the star than $15^{''}$ \footnote{vdB 133 shows the  
3.3 $\micron$ emission in the nebulosity located beyond $15^{''}$ from the exciting star.}.  
This circumstellar disk dust will have similar dilution factors from star to star so
T$_{eff}$ can be used to estimate the relative strength of the UV radiation field. But it can be seen from Figure 3 
that T$_{eff}$ is no better correlated with the presence of the 3.3 $\micron$ feature than with $G_{0}$. 
So neither T$_{eff}$ nor $G_{0}$ can be used to predict whether PAH emission will be seen along a particular sightline. 
One clue may come from the fact that 5 of the 6 stars that show PAH emission in the Meeus sample also have IR excesses 
indicating the presence of warm dust. So geometry may be important. Portions of the disk which are warm will probably 
correspond to photo-dissociation regions that are receiving photons from the star. So we may be viewing these
disks more pole-on. Whereas, thick disks viewed edge-on will be less likely to show an IR excess or PAH emission
because we are seeing only cold dust that is not being illuminated by any stellar photons. Flared disks tend to have 
UIBs whereas non-flared disks have weak to no UIB detections (Meeus et al. 2001). Perhaps the flaring of these 
disks increases the viewing angle where warm dust can be seen. 

Since cool stars with small UV radiation fields such as HD 34700 show 3.3 $\mu$m emission, the trend in Figure 3 
supports the prediction of the astronomical PAH model of Li \& Draine (2002), that the 3.3 $\micron$ UIB 
should be observable in UV-weak objects.  But Figure 3 also shows that relatively hot stars with high UV radiation 
fields do not show any such emission.  So our results suggest that the strength of the UV radiation field impinging 
on the dust is not the only or even the most important factor in determining whether the 3.3 $\micron$ UIB is present in a 
dusty astrophysical object. In addition to simple disk-viewing angle considerations, electron density variations from 
star to star may be an important factor in whether the 3.3 $\mu$m emission feature is observable in an object.  However, 
a full spectrum of each star extending from 3.3 to 10 $\mu$m is needed in order to solidly model the UIB features 
and estimate the electron density.  It is also likely that the ionization state of the PAHs in each object will 
affect the observability of the 3.3 $\mu$m emission feature.  Laboratory 
studies (Allamandola et al. 1999) show the 3.3 $\mu$m absorption feature to be very strong, relative to the features 
in the 6 to 10 $\mu$m region, for a mixture of neutral PAHs, and to be much weaker, with respect to features in the 
6 to 10 $\mu$m, for the same mixture of PAHs in their positive state.  We conclude that the task of determining 
the role of the impinging radiation field on the observability and variations in the 3.3 $\mu$m emission feature in 
our targets would be furthered by spectra from the same stars covering the 6-10 $\mu$m region.

\acknowledgements

T.L. Smith, G.C Clayton and L. Valencic were visiting astronomers at the Infrared Telescope Facility, which 
is operated by the University of Hawaii under Cooperative Agreement no. NCC 5-538 with the National 
Aeronautics and Space Administration, Office of Space Science, Planetary Astronomy Program.  Aigen Li is 
thanked for his interest and contribution to Figure 1, the model results from Li \& Draine (2002).  We also thank 
K. Uchida for the the ISOCAM data points to the same figure.  We would like to gratefully acknowledge 
G. Meeus for the helpful discussion on the Herbig Ae/Be objects included in this study, and B. 
Vandenbussche for the ISO-SWS spectrum provided in Figure 2.  Finally, we would like to thank A. G. G. M. 
Tielens for his helpful comments on this paper.

\clearpage

\begin{figure}[th]
\begin{center}
\epsscale{0.75}
\plotone{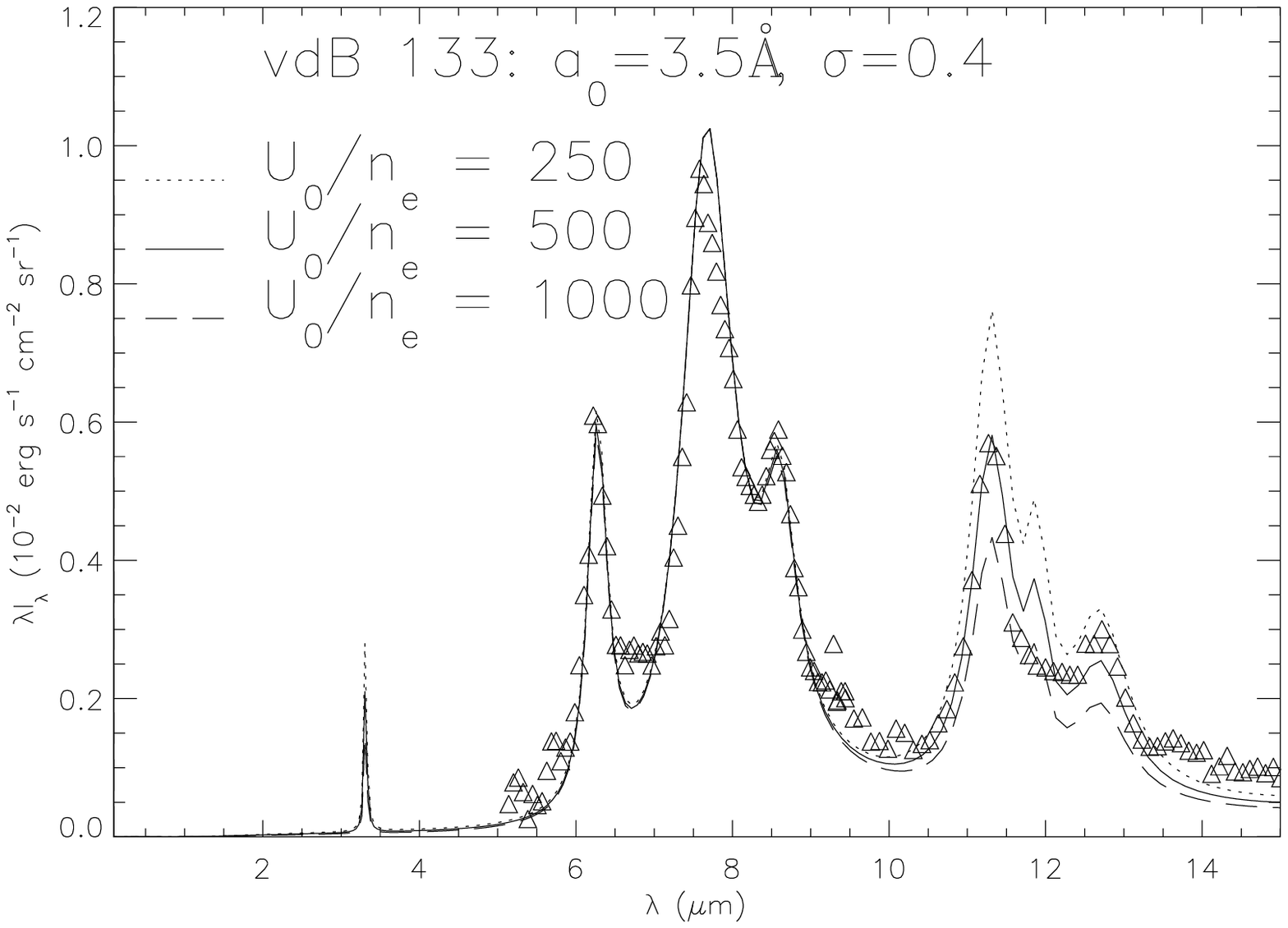}
\end{center}
\caption{The model-predicted IR emission spectra for PAHs in the moderate UV environment of 
vdB 133 in comparison with the ISOCAM (triangles) spectrum (Li \& Draine 2002, Uchida et al. 1998).  
A lognormal size distribution characterized by three parameters, a$_{0}$ and $\sigma$ and b$_{C}$ was adopted 
which determine the peak location, the width of the lognormal distribution, and the total amount of 
C atoms (relative to H) in the PAHs.  The calculations were done for three ratios of U$_{0}$ (the ratio 
of the 912 \AA - 1$\mu$m energy density relative to the value for the Mathis, Merzer, \& Panagia (1983) 
solar neighborhood interstellar radiation field) to electron density.}
\end{figure}

\begin{figure}[th]
\begin{center}
\epsscale{1.1}
\plotone{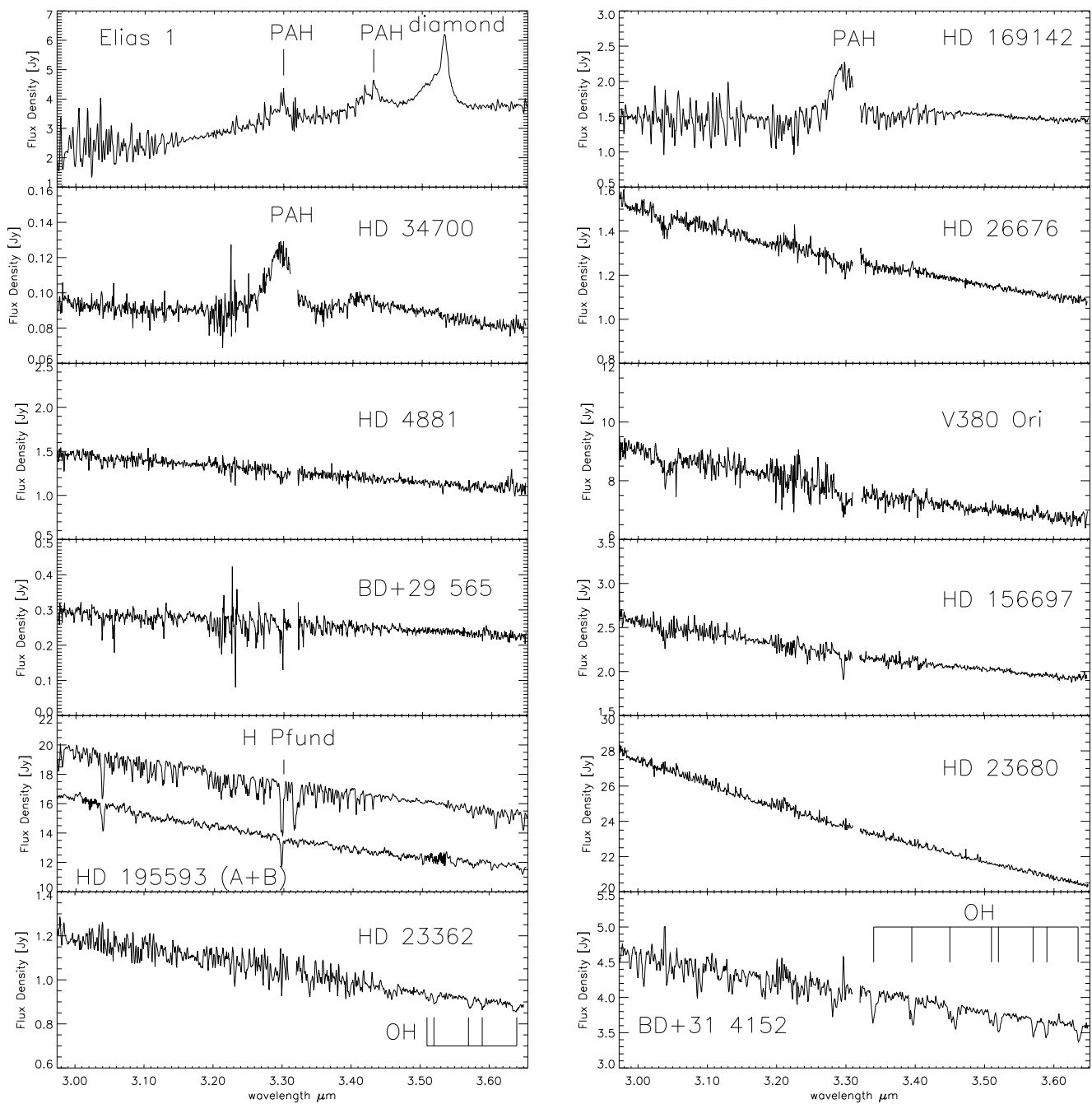}
\end{center}
\caption{The 3.3 $\micron$ spectral region in six stars embedded in reflection 
nebulae, and in six Vega-like stars observed with SpeX at the NASA Infrared Telescope 
Facility (IRTF) in Aug./Sept. 2002.  A spectrum of HD 195593 (A+B) from the Vandenbussche et al. (2002) 
near-infrared stellar spectra atlas collected with the ISO-SWS is included with our observations, plotted 
as the bottom spectrum.  A gap has been left in the spectrum where imperfect telluric removal resulted in 
spurious data.}
\end{figure}

\begin{figure}[th]
\begin{center}
\epsscale{1.}
\plotone{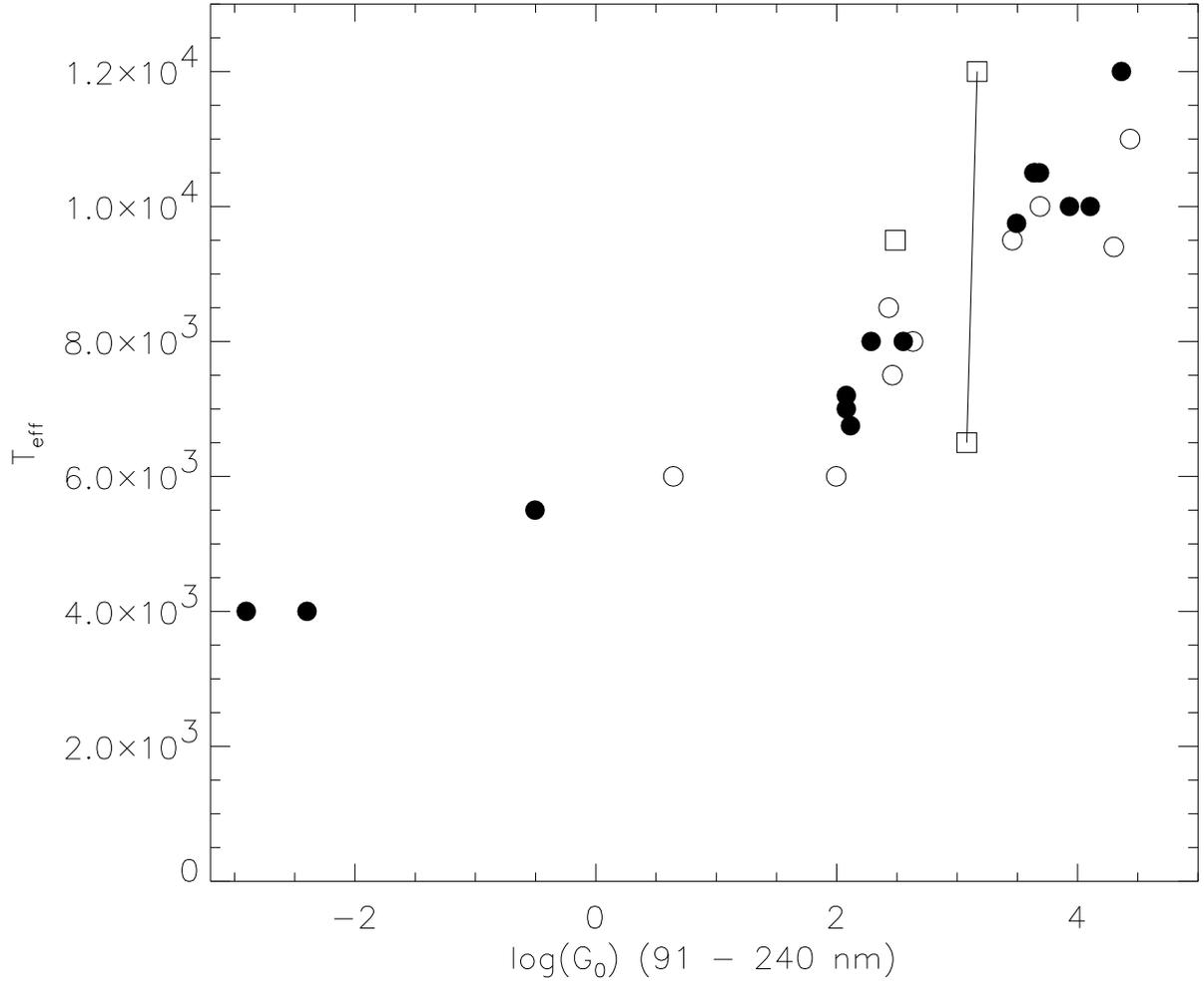} 
\end{center}
\caption{The log of the UV radiation field, G$_{0}$, estimated at 15$^{''}$ NS offset for each star, 
versus $T_{eff}$, where G$_{0}$ is in units of 1.6 x 10$^{-3}$ erg $s^{-1} cm^{-2}$, 
the local interstellar radiation field according to Habing (1968) and used by Uchida et al. (2000).  
The 3.3 $\micron$ feature detections are plotted with open circles, non-detections with filled circles.  HD 195593 
(A+B) (connected with a solid line) and V380 Ori have been plotted with open squares. Emission is
seen in these objects at further than 15\arcsec~from the stars. See text.}
\end{figure}

\clearpage
\begin{deluxetable}{cccc}
\tabletypesize{\scriptsize}
\tablecaption{IRTF-observed Vega-like stars and stars embedded in reflection nebulae. \label{tbl-1}}
\tablewidth{0pt}
\tablehead{
\colhead{Star} & \colhead{Spectral Type}   & \colhead{Dust Type}   &
\colhead{UIBs} 
}
\startdata
HD 26676 &B8 Vn &Vega-like &\\
HD 4881 &B9.5 V &Vega-like&\\
V380 Ori &A0-2 &Reflection Nebula&\\
Elias 1 &A0-6 &Reflection Nebula&3.3, 6.2, 7.7, 8.6 \& 11.2\\
HD 169142 &A5 Ve  &Vega-like&3.3, 6.2, 7.7, 8.6 \& 11.3\\
BD+29 565 &F0 V &Reflection Nebula&\\
HD 156697 &F0 n&Reflection Nebula&\\
HD 195593 (A+B) & FV Iab + B7 II&Reflection Nebula&3.3, 6.2, 7.7, 8.6, 11.3, 12.7 \& 5-15 micron continuum emission\\
HD 34700 &G0V&Vega-like&3.3, 6.2, 7.7 \& 8.6\\
HD 23680 &G5&Vega-like&\\
HD 23362 &K2&Vega-like&\\
BD+31 4152 &M1 IIIe&Reflection Nebula&\\
\enddata
\end{deluxetable}

\begin{deluxetable}{cccccccc}
\tabletypesize{\scriptsize}
\tablecaption{Estimates of $G_{0}$ at a stellar offset of $15^{''}$. \label{tbl-2}}
\tablewidth{0pt}
\tablehead{
\colhead{Star\tablenotemark{*}} & \colhead{$T_{eff}$ (K)} & \colhead{Nebula Type} & \colhead{Star Distance (pc)}   & 
\colhead{$G_{0}$ (91-240 nm)} & \colhead{100 $\mu$m sky\tablenotemark{a}} & \colhead{disk} & 
\colhead{References\tablenotemark{b}} 
}
\startdata
HD 100546* &11000 &Herbig Ae/Be &103 &27288 &40 &yes &1\\
HD 26676 &12000 &Vega-like &151  &23125 &42 &yes &2\\
HD 20041* &9400  &Reflection Nebula &1400 &20000 & &yes &6, 10\\
HD 158643 &10000 &Herbig Ae/Be &131 &12725 &140 &yes &1\\
HD 150193 &10000 &Herbig Ae/Be &150 &8563  &143 &yes &1\\
HD 179218* &10000 &Herbig Ae/Be &240 &4875  &113 &yes &1\\
HD 4881 &10500 &Vega-like &168 &4829 & &yes &2\\
HD 104237 &10500 &Herbig Ae/Be &116 &4438 &22 &yes &1\\
HD 163296 &10500 &Herbig Ae/Be &122 &4331 & &yes &1\\
HD 31293 &9750 &Herbig Ae/Be &144 &3113 &45 &yes &1\\
Elias 1* &9500 &Herbig Ae/Be &150 &2871 & 43 &yes &5, 6, 7\\
HD 195593 (A) &6500 &Reflection Nebula &1500 &1202 & &no &6\\
HD 195593 (B) &12000 &Reflection Nebula &1500 &1466 & &no &6\\
HD 169142* &8000 &Herbig Ae/Be &145 &430 &58 &yes &1\\
HD 139614 &8000 &Herbig Ae/Be &151 &357 &52 &yes &1\\
V380 Ori &9500 &Vega-like &460 &307 &193 &yes &4, 6\\
HD 100453* &7500 &Herbig Ae/Be &115 &290 &30 &yes &1\\
HD 142666* &8500 &Herbig Ae/Be &200 &270 &61 &yes &1\\
HD 144432 &8000 &Herbig Ae/Be &200 &193 &67 &yes &1\\
HD 135344 &6750 &Herbig Ae/Be &84  &130 &28 &yes &1\\
BD+29 565 &7200 &Reflection Nebula &140 &120 & &no &6, 8\\
HD 156697 &7000 &Reflection Nebula &139 &120 & &no &6, 9\\
HD 142527* &6000 &Herbig Ae/Be &200 &99 &88 &yes &1\\
HD 34700* &6000 &Vega-like &180 &4.4 & 28 &yes &3\\
HD 23680 &5500 &Vega-like &205 &0.31 & &yes &2\\
BD+31 4152 &4000 &Reflection Nebula &350 &0.004 & &no &6, 8\\
HD 23362 &4000 &Vega-like &187 &0.001 & 23 &yes &2\\
\enddata
\tablenotetext{*}{Objects with a positive 3.3 $\mu$m UIB detection are noted with a *.}
\tablenotetext{a}{(1) Meeus et al. (2001), (2) Kalas et al. (2002), (3) Walker \& Heinrichsen (2000), 
(4) Rossi et al. (1999), (5) Allen et al. (1982), (6) Sellgren et al. (1996), (7) Van Kerckhoven et al. 
(2002), (8) Ramesh (1994), (9) Antonello \& Montegazza (1997), (10) Hovhannessian \& Hovhannessian (2001).}
\tablenotetext{b}{The total IRAS brightness of the sky surrounding the source in a 0.5 degree beam at 100 $\mu$m, 
in MJy/Sr.}
\end{deluxetable}


\clearpage

\begin{thebibliography}{}
\bibitem[ALLAMANDOLA et al.(1985)]{ala85} Allamandola, L. \ J. \, Tielens, A. \ G. \ G. \ M. \, \& Barker, 
J. \ R. \ 1985, \apj, 290 l25
\bibitem[Allamandola et al. (1989)]{ala89} Allamandola, L. \ J. \, Tielens, A. \ G. \ G. \ M. \& Barker, 
J. \ R. \ 1989, \apjs, 71, 733
\bibitem[Allamandola et al. (1999)]{ala99} Allamandola, L. \ J. \, Hudgins, D. \ M. \& Sandford, S. \ A. \ 
1999, \apj, 511, L115
\bibitem[Allen et al. (1982)]{allen82} Allen, D. \ A. \, Baines, D. \ W. \ T. \, Blades, J. \ C. \& 
Whittet, D. \ C. \ B. \ 1982, \mnras, 199, 1017
\bibitem[Antonello (1997)]{ant97} Antonello, E. \& Montegazza L. \ 1997, \aap, 327, 240
\bibitem[Bakes et al. (1994)]{bakes94} Bakes, E. \ L. \ O. \& Tielens, A. \ G. \ G. \ M. \ 1994, \apj, 
427, 822
\bibitem[Bakes et al. (1998)]{bakes98} Bakes, E. \ L. \ O. \& Tielens, A. \ G. \ G. \ M. \ 1998, \apj, 
499, 258
\bibitem[Bakes et al. (2001)]{bakes01} Bakes, E. \ L. \ O. \, Tielens, A. \ G. \ G. \ M. \& Bauschlicher, 
C. \ W. \ 2001, \apj, 556, 501
\bibitem[Clayton et al. (2003)]{clay03} Clayton, G. \ C. et al. 2003, \apj, 592, 947
\bibitem[Cohen et al. (1975)]{cohen75} Cohen et al. 1975, \apj, 196, 179
\bibitem[Cook \& Sakally (1998)]{cooksak98} Cook, D. \ J. \& Sakally, R. \ J. \ 1998, \apj, 493, 793
\bibitem[Cushing et al. (2003)]{cush03} Cushing, M. \ C. \, Vacca, W. \ D. \, \& Rayner, J. \ T. \ 2003,
\pasp, 999, 999
\bibitem[Dartois et al. (1997)]{dartois97} Dartois, E. \& d'Hendecourt, L. \ 1997, \aap, 323, 534
\bibitem[Duley (1981)]{duley81} Duley, W. \ W. \& Williams, D. \ A. \ 1981, \mnras, 241, 697
\bibitem[Elias (1978)]{elias78} Elias, J. \ H. \ 1978, \aap, 70, L3
\bibitem[Geballe et al. (1989b)]{gebal89b} Geballe, T. \ R. \, Tielens, A. \ G. \ G. \ M. \, Allamandola, 
L. \ J. \, Moorhouse, A. \& Brand, P. \ W. \ J. \ L. \ 1989b, \apj, 341, 278
\bibitem[Geballe et al. (1992)]{gebal92} Geballe, T. \ R. \, Tielens, A. \ G. \ G. \ M. \, Kwok, S.\& 
Hrivnak, B. \ J. \ 1992, \apj, 387, L89
\bibitem[Geballe et al. (1997)]{gebal97} Geballe, T. \ R. \ 1997, in ASP Conf. Ser. 122, From Stardust to 
Planetesimals, ed. Y. \ J. \ Pendleton \& A. \ G. \ G. \ M. \ Tielens (San Francisco: ASP), 119
\bibitem[Genzel \& Cesarsky (2000)]{gences00} Genzel, R. \& Cesarsky, C. \ J. \ 2000, Annu. Rev. Astron. 
Astrophys., 38, 761 
\bibitem[Gillett et al. (1973)]{gillet73} Gillett, F. \ C. \, Forrest, W. \ J. \, \& Merrill, K. \ M. \ 
1973, \apj, 183, 87
\bibitem[Guillois et al. (1999)]{guil99} Guillois, O. \, Ledoux, G. \& Reynaud C. \ 1999, \apj, 521, 133
\bibitem[Habing (1968)]{habing68} Habing, H. \ J. \ 1968, Bull. Astron. Inst. Netherlands, 19, 421
\bibitem[Hovhannessian \& Hovhannessian (2001)]{hovhov01} Hovhannessian, R. \ Kh. \& Hovhannessian, E. \
R. \ 2001, Astrophysics, 44, 454
\bibitem[Kalas et al. (2002)]{kalas02} Kalas, P. \, Graham, J. \ R. \, Beckwith, S. \ V. \ W. \, 
Jewitt, D. \ C. \& Lloyd, J. \ P. \ 2002, \apj, 567, 999
\bibitem[Kraemer (2002)]{kraemer} Kraemer, G. \ C. \, Sloan, G. \ C. \, Price, S. \ D. \& Walker, H. \ 
J. \ 2002, \apjs, 140, 389 
\bibitem[Kurucz (1992)]{kurucz92} Kurucz, R. \ L. \ 1992,  in IAU Symposium 149, The Stellar Populations 
of Galaxies, ed. B. \ Barbuy \& A. \ Renzini (Dordrecht:Kluwer), 225 
\bibitem[Leger \& Puget (1984)]{lejpug84} Leger, A. \, \& Puget, J. \ L. \ 1984, \aap, 137, L5
\bibitem[Lemke et al. (1998)]{lemke98} Lemke, D. \ et al. \ 1998, \aap, 331, 742
\bibitem[Li \& Draine (2001)]{lidran01} Li, A. \& Draine, B. \ T. \ 2001, \apj, 550, L213
\bibitem[Li \& Draine (2002)]{lidrain02} Li, A. \, \& Draine, B. \ T. \ 2002, \apj, 572, 232
\bibitem[Mathis et al. (1983)]{mathis83} Mathis, J. \ S. \,  Mezger, P. \ G. \& Panagia, N \ 1983, 
\aap, 128, 212
\bibitem[Matilla et al. (1996)]{mat96} Matilla, K. \ et al. \ 1996, \aap, 315, L353
\bibitem[Meeus et al. (2001)]{meeus01} Meeus, G. \ et al. 2001, \aap, 365, 476
\bibitem[Merrill et al. (2000)]{merrill00} Merrill, K. \ M. \, Soifer, B. \ T. \,\& Russell, R. \ W. \ 
1975, \apj, 2000, L37
\bibitem[Ramesh (1994)]{ramesh94} Ramesh, B. \ 1994, J. Astrophys. Astr., 15, 415
\bibitem[Rayner et al. (2003)]{ray03} Rayner J. \ T. \, Toomey D. \ W. \, Onaka, P. \ M. \, Denault, A. \ J. \, 
Stahlberger, W. \ E. \, Vacca, W. \ D. \, Cushing, M. \ C. \, \& Wang, S. \ 2003,
\pasp, 155, 362
\bibitem[Roche (1989)]{roche89} Roche, P. \ F. \ 1989, IAU Symp. 135, Interstellar Dust ed L. J. Allamandola 
and A. G. G. M. Tielens (Dordrecht:Kluwer), 303
\bibitem[Roelfsema (1996)]{roelf} Roelfsema, P. \ R. \ et al. 1996, \aap, 315, L289
\bibitem[Rossi et al. (1999)]{rossi99} Rossi, C. \ et al. \ 1999, \aap sup., 136, 95
\bibitem[Russell et al. (1977)]{rus77} Russell, R. \ W. \, Soifer, B. \ T. \, \& Merrill, K. \ M. \ 1977, 
\apj, 213, 66
\bibitem[Salama et al. (1996)]{salam96} Salama, F. \, Bakes, E. \ L. \ O. \, Allamandola, L. \ J. \& 
Tielens, A. \ G. \ G. \ M. \ 1996, \apj, 458, 621 
\bibitem[Schutte et al. (1993)]{schut93} Schutte, W. \ A. \, Tielens, A. \ G. \ G. \ M. \& Allamandola, 
L. \ J. \ 1993, \apj, 415, 397
\bibitem[Sellgren et al. (1996)]{selg96} Sellgren, K. \, Werner, M. \ W. \& Allamandola, L. \ J. \ 1996, 
\apjs, 102, 369
\bibitem[Sellgren et al. (1990)]{selg90} Sellgren, K. \, Luan, L.\, \& Werner, M. \ W. \ 1990, 
\apj, 359, 384
\bibitem[Sellgren et al. (1990b)]{selg90b} Sellgren, K. \, Tokunaga, A. \ T. \& Nakada, Y. \ 1990, \apj, 
349, 120
\bibitem[Sellgren et al. (2001)]{sel01} Sellgren, K. \ 2001, Spectrochimica Acta Part A: Molecular 
and Biomolecular Spectroscopy, 57, 627
\bibitem[Soifer et al. (1976)]{soif76} Soifer, B. \ T. \, Russel, R. \ W. \, \& Merrill, K. \ M. \ 1976, \apj, 
250, 631
\bibitem[Tielens (1993)]{tielens93} Tielens, A. \ G. \ G. \ M. \ 1993, Dust and Chemistry in Astronomy 
ed T. \ J. \ Millar and D. \ A. \ Williams (Bristol: IOP Publishing), 103
\bibitem[Tokunaga (1997)]{tok97} Tokunaga, A. \ 1997, in ASP Conf. Ser. 124, Diffuse Infrared Radiation and 
the IRTS, ed. H. \ Okuda, T. \ Matsumoto, \& T. \ L. \ Roellig (San Francisco: ASP), 149
\bibitem[Uchida et al. (1988)]{uchica88} Uchida, K. \ I. \, Sellgren, K.\ \& Werner, M. \ W. \
1998, \apj, 493, L109
\bibitem[Uchida et al. (2000)]{uchida00} Uchida, K. \ I. \, Sellgren, K. \, \& Werner, M. \ W. \& Houdashelt M. \ L. \ 2000, 
\apj, 530, 817
\bibitem[Vacca et al. (2003)]{vaca03} Vacca, W. \ D. \, Cushing, M. \ C. \, \& Rayner, J. \ T. \ 2003,
\pasp,155, 389
\bibitem[Van Kerckhoven et al. (2002)]{vankerc02} Van Kerckhoven, C. \, Tielens, A. \ G. \ G. \ M. \, \& 
Waelkens, C. \ 2002, \aap, 384, 568
\bibitem[Verstraete et al. (2001)]{verst01} Verstraete, L. \ et al. 2001, \aap, 372, 981
\bibitem[Walker \& Heinrichsen (2000)]{walkhein00} Walker, H. \ J. \, \& Heinrichsen, I. \ 2000, 
ICARUS, 143, 147
\end{thebibliography}
\end{document}